\documentclass[10pt,conference]{IEEEtran}
\IEEEoverridecommandlockouts
\usepackage{cite}
\usepackage{amsmath,amssymb,amsfonts}
\usepackage{algorithmic}
\usepackage{graphicx}
\usepackage{textcomp}
\usepackage{xcolor}
\usepackage{hhline}
\usepackage{multirow}
\usepackage{tabularx}
\usepackage{booktabs}
\usepackage{xurl}
\usepackage{hyperref}
\def\BibTeX{{\rm B\kern-.05em{\sc i\kern-.025em b}\kern-.08em
    T\kern-.1667em\lower.7ex\hbox{E}\kern-.125emX}}
\begin{document}

\title{From Hazard Identification to Controller Design: Proactive and LLM-Supported Safety Engineering for ML-Powered Systems
}
\newcommand{\todo}[1]{\textit{\textcolor{green}{#1}}}

\author{
    \IEEEauthorblockN{Yining Hong}
    \IEEEauthorblockA{
        \textit{Carnegie Mellon University}\\
    }
    \and
    \IEEEauthorblockN{Christopher S. Timperley}
    \IEEEauthorblockA{
        \textit{Carnegie Mellon University}\\
    }
    \and
    \IEEEauthorblockN{Christian K{\"a}stner}
    \IEEEauthorblockA{
        \textit{Carnegie Mellon University}\\
    }
}

\maketitle

\begin{abstract}
Machine learning (ML) components are increasingly integrated into software products, yet their complexity and inherent uncertainty often lead to unintended and hazardous consequences, both for individuals and society at large. Despite these risks, practitioners seldom adopt proactive approaches to anticipate and mitigate hazards before they occur. Traditional safety engineering approaches, such as Failure Mode and Effects Analysis (FMEA) and System Theoretic Process Analysis (STPA), offer systematic frameworks for early risk identification but are rarely adopted. This position paper advocates for integrating hazard analysis into the development of any ML-powered software product and calls for greater support to make this process accessible to developers. By using large language models (LLMs) to partially automate a modified STPA process with human oversight at critical steps, we expect to address two key challenges: the heavy dependency on highly experienced safety engineering experts, and the time-consuming, labor-intensive nature of traditional hazard analysis, which often impedes its integration into real-world development workflows. We illustrate our approach with a running example, demonstrating that many seemingly unanticipated issues can, in fact, be anticipated.
\end{abstract}


\section{Introduction}

Phrases like \textit{``this was an unintended consequence''} or \textit{``nobody could have anticipated this new problem''} often arise when software products face issues, such as mistakes or biases in ML models within the software. For example, applications may unintentionally amplify biases\cite{bias_amplification_suresh2021framework} or introduce privacy risks\cite{privacy_risks_chen2021machine}. 
However, \textit{unintended consequences}  merely implies that these issues were not anticipated, not that they were inherently unpredictable; 
had they been anticipated, developers could, in many cases, have mitigated them before they occurred.
This position paper argues that (1) systematic \textit{hazard analysis} methods from safety engineering are well suited to anticipate a wide range of harms in ML-powered applications beyond traditional safety risks, and (2) LLM-supported automation can make hazard analysis more accessible and manageable in terms of effort. We believe that hazard analysis is particularly effective in the early development stages, helping identify and design controllers to mitigate harm.


Recent research has explored more or less structured strategies to anticipate potential harms, primarily bias and fairness issues, in the context of ML impact assessment\cite{pagano2022bias,mehrabi2021survey,buccinca2023aha,wang2024farsight, microsoft_rai_template_2022}. However, these approaches are generally model-centric and applied ad-hoc \cite{jatho2022system}, overlooking controllers beyond the model such as safeguards, trend monitoring, human oversight, and user interface modifications -- areas where software engineers can contribute to responsible engineering of ML-powered systems, extending beyond model-focused considerations.

More recent research has adopted traditional safety engineering, especially hazard analysis methods such as \textit{Failure Mode and Effects Analysis (FMEA)} and \textit{System Theoretic Process Analysis (STPA)}, to proactively identify a broad range of ethical and social risks in ML models and ML-powered software systems\cite{khlaaf2022hazard, Dobbe_2022, martelaro2022exploring, rismani2023beyond, rismani2023plane, jatho2022system, rismani2024silos, rismani2021ai}. 
However, outside traditional safety-critical domains such as autonomous vehicles\cite{adler2016safety, abdulkhaleq2017using}, existing studies largely discuss only the \textit{potential} application of hazard analysis, with several emerging challenges:
First, traditional hazard analysis methods are limited by predefined system boundaries, restricting the consideration of a broader scope of risks and solutions.
Second, these methods require substantial time and effort, making them costly and challenging to integrate into fast-paced, continuous development workflows outside traditional safety-critical systems\cite{rismani2023beyond, martelaro2022exploring}.
Third, the responsibility for identifying and mitigating risks often falls to software developers or ML model creators, who may lack expertise in safety engineering, potentially reducing the effectiveness and thoroughness of these analyses\cite{martelaro2022exploring}.

In this paper, we provide a concrete illustration of how hazard analysis, with minor modifications, can proactively anticipate harms and, more importantly, guide the design of controllers to mitigate those harms before they occur at both the model and system levels. Furthermore, we show how LLMs can support this process by offering guidance to software engineers and data scientists with limited safety-engineering expertise, requiring only moderate efforts that align with the fast-paced development practices of ML-powered applications.

In summary, this position paper offers a fresh and critical perspective on using hazard analysis to broadly anticipate harms in ML-powered applications and to design system-level controllers, illustrated with a concrete example. We further contribute a discussion and demonstration of the potential of how LLMs can support developers in the hazard analysis process. 
We envision that a lightweight, LLM-supported hazard analysis process will become a routine step in the responsible engineering of ML-powered applications, enabling the mitigation of many harms well before they occur.


\begin{figure*}[th]
\centerline{\includegraphics[width=0.95\linewidth]{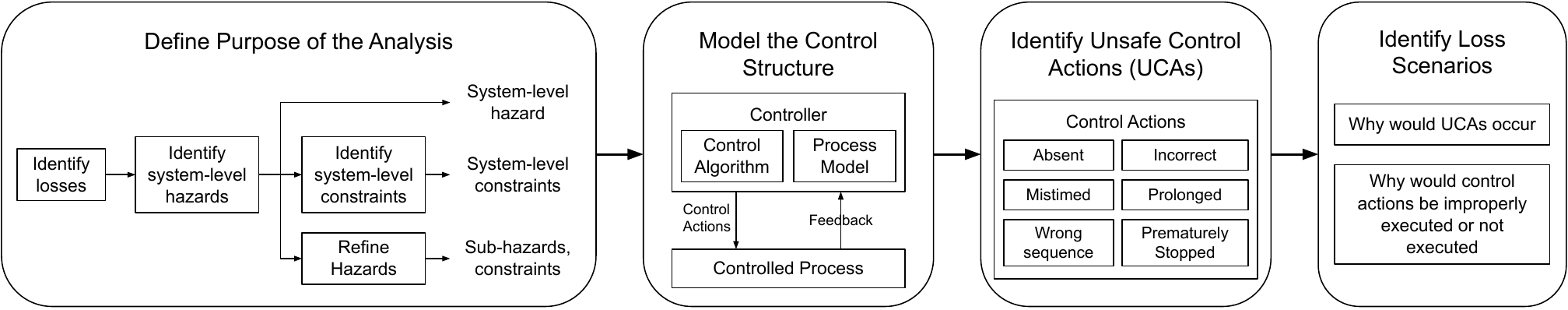}}
\caption{An overview of the STPA method, based on the textual description and figures in the STPA handbook\cite{STPA_Handbook_Stpa2018}.}
\label{figure: STPA method overview}
\end{figure*} 

\section{Scope and Related Work}\label{section: Related Works}

Though there are various definitions for AI and ML systems, in this paper, we use the term ``ML-powered system'' to denote any software system that incorporates ML models as a component, aiming to emphasize the system as a whole with ML model as an inherently `unreliable' module within it.

Practitioners and researchers have acknowledged the potential social risks posed by ML-powered systems\cite{bender2021dangers, weidinger2021ethical}, such as bias and insufficient oversight.
Technical approaches typically focus on measuring and addressing model-level issues, focusing on aspects such as fairness, shortcut reasoning, or privacy\cite{zemel2013learning, shokri2015privacy, bias_amplification_suresh2021framework, privacy_risks_chen2021machine}, often neglecting broader system or environmental considerations.
Recent research has introduced structured approaches and tools to help practitioners anticipate potential harms of ML-powered systems, ranging from impact assessment templates\cite{MicrosoftAITools} to tools that facilitate more or less structured brainstroming\cite{buccinca2023aha, wang2024farsight, kieslich2023anticipating, bogucka2024co}. While these methods effectively identify relatively explicit harms or hazards linked to the system’s overarching purpose, they fall short in analyzing the detailed components or control structures within these systems and are often used in an ad-hoc manner\cite{martelaro2022exploring}.

Safety engineering has a long history \cite{bahr2014system}, with traditional methods having been applied in various safety-critical domains, such as aviation systems\cite{leveson2004new, SAE_ARP4761A_2023} and autonomous vehicles \cite{adler2016safety, abdulkhaleq2017using}.
Recent research has applied these safety engineering methods, particularly the hazard analysis frameworks FMEA\cite{rismani2023plane, rismani2021ai} and STPA\cite{rismani2023beyond, rismani2023plane, jatho2022system, rismani2024silos}, to identify potential hazards and assess control structures in ML-powered systems. Findings suggest that such systematic approaches are not only effective in identifying a wider spectrum of hazards but are also potentially applicable throughout all development stages.
However, integrating safety engineering frameworks into the development cycles of ML systems is often impractical due to the extensive time and paperwork required \cite{rismani2023beyond, martelaro2022exploring}. Many practitioners lack in-depth safety engineering expertise, which can reduce the effectiveness of these frameworks \cite{martelaro2022exploring}. Furthermore, most existing studies still focus on the model and its development process, with limited attention to non-ML components and the social environment.
This paper demonstrates that hazard analysis is effective in anticipating a wide range of hazards and aids in proactive controller design. Furthermore, it shows that LLMs are promising in supporting humans conducting STPA by reducing effort and encouraging broader, more comprehensive thinking beyond the model.

\section{A Brief Introduction to STPA} \label{section: Standard STPA}

While safety engineering techniques date back to the 1950s\cite{safeware_10.1145/202709, safety_critical_computer_systems_10.5555/524721}, our approach builds on System-Theoretic Process Analysis (STPA), a state-of-the-art hazard analysis framework grounded in the System-Theoretic Accident Model and Processes (STAMP) \cite{Engineering_a_safer_world_Nancy2012}. We use STPA because it is designed for analyzing highly complex systems, can be applied from the early development stages, and considers both technical and human factors \cite{Engineering_a_safer_world_Nancy2012} -- all crucial for ML-powered software. 

Traditionally, STPA follows four steps, outlined in 
Figure~\ref{figure: STPA method overview}:

1)\textit{ Define the Purpose of the Analysis:} 
We list \textit{stakeholders} and their values, then identify important \textbf{losses} (e.g., loss of life). Based on the losses, we determine system-level \textbf{hazards}, which are conditions that may lead to a loss (e.g., aircraft too close to other objects). Hazards are then inverted into \textbf{constraints}, which are system conditions that need to be satisfied to prevent hazards. This way, STPA anticipates harms and establishes safety requirements to mitigate them.

2)\textit{ Model the Control Structure: } 
We outline the control structure designed to ensure the safety constraints are met. Each previously identified constraint targeted for resolution should have at least one associated \textbf{controller} (e.g., an automated safeguard or a human supervisor). New controllers can be envisioned for constraints without sufficient controllers. STPA particularly focuses on feedback-control loops between controllers and the controlled processes. 

3)\textit{ Identify \textbf{Unsafe Control Actions}: } 
We analyze each controller's potential failure modes by considering whether a hazard may arise if each control action is absent, incorrect, mistimed, executed in the wrong sequence, prolonged, or stopped prematurely. Control action errors that can lead to unsafe outcomes are then analyzed in the final step.

4)\textit{ Identify Loss Scenarios: } 
We examine why unsafe control actions may occur, potentially leading to revising existing controllers or introducing additional controllers (e.g., pilots may be unreliable in distance checking, hence we may introduce an automated warning system). The STPA process is then repeated with the modified control structure.

In a nutshell, STPA is a structured approach that works backward from harms (losses) to safety requirements (constraints) and then to mitigation strategies (controllers) and their potential failures. 
It encourages broad consideration of the control structure, including non-technical controllers like training, human oversight, and government regulations.

Throughout the paper, we use safety engineering terminology for illustrations and discussions. Since this may be unfamiliar to most software engineers and data scientists, a summary and mapping are provided in Table~\ref{tab:terms}.

\begin{table}
    \centering
        \caption{Terminology of safety engineering and mapping to concepts in ML-powered application development}
    \label{tab:terms}
    \begin{tabular}{p{1.05cm}p{6.8cm}}
    \toprule
    \textbf{Concept} & \textbf{Explanation and mapping to ML-powered application}\\\midrule
        Loss & An event undesirable to a stakeholder, including bodily harm, allocation harms (e.g., discriminatory service), representation harms, and societal harms (e.g., polarization)  \\ \addlinespace[2pt]
        Hazard & A state or a set of conditions that may lead to a loss in certain scenarios, such as when an ML-powered software system produces faulty information \\\addlinespace[2pt]
        Constraint & A {(safety) requirement} that, when satisfied, prevents the hazard and thus the loss (e.g., an ML-powered system does not produce faulty information or warns the user if it does) \\\addlinespace[2pt]
        Controller & A technical or non-technical components that enforce constraints, including technical safeguards (e.g., exception handling, redundant sensors, I/O validation) and human oversight (e.g., processes, training, monitoring, alerting) \\\addlinespace[2pt]
        Unsafe Control Action & 
        A control action, or lack thereof, provided by a controller can, under certain conditions, lead to a hazard (e.g., output validation suppressing correct answers) \\
         \bottomrule
    \end{tabular}
\end{table}

STPA is a labor-intensive process needing expert involvement and is mainly used for safety-critical systems like aviation, focusing on severe losses such as loss of life. Researchers found STPA potentially useful, but its time-consuming nature is unsuitable for rapid development, and developers often lack the necessary expertise \cite{rismani2023beyond, martelaro2022exploring, rismani2023plane}.

\section{LLM-Supported Hazard Analysis to Anticipate Harms in ML-Powered Applications} \label{section: Method}

We argue that hazard analysis is effective in anticipating a wide range of harms in ML-powered applications, such as fairness, usability issues, and societal concerns like deskilling and polarization, extending beyond the traditional focus on severe harms such as loss of life in safety-critical systems.
Specifically, hazard analysis aids in \textit{anticipating} problems and \textit{designing} mitigation strategies in systems that might otherwise be released without any.
In this section, we walk through the four STPA steps using a running example of a simple web system designed to recommend outdoor trails.

Evidence suggests that routinely applying hazard analysis to everyday software applications as part of responsible engineering practices is challenging due to high costs and skill requirements. Therefore, we also present how LLMs, specifically GPT-4o by OpenAI \cite{GPT4_openai2023gpt4}, can assist software engineers and data scientists in conducting such analysis without requiring extensive training and excessive paperwork.


Our running example is inspired by the existing customized GPT assistant \textit{AllTrails} \cite{chatgpt_alltrails}. The application operates within the ChatGPT chat interface and is guided by system prompts from developers. It accesses the predefined \url{alltrails.com} API to fetch trail information. Upon receiving user prompts (e.g., ``Recommend trails near [district name] that are dog-friendly.''), it calls the API for information and provides a response. To our knowledge, it lacks additional controllers beyond its system prompt instructions and ChatGPT's built-in safeguards.
As is often the case, the LLM should be considered an unreliable component, as it may fail to follow instructions, hallucinate, or produce factual errors. We assume this application will be extended beyond its current form and expect it to gain significant popularity. Note that we explicitly chose a running example of an everyday ML-powered application, rather than a traditional safety-critical system, to illustrate how hazard analysis can proactively anticipate issues worth mitigating, even for such seemingly harmless applications.

{






}

\subsection{Anticipating Losses and Hazards}
\textit{Stakeholders:} Following traditional STPA, we begin with identifying stakeholders and their values to assess system losses. To go beyond identifying only the most severe losses, we encourage a comprehensive exploration of stakeholders, including indirect ones, which extends the scope typically practiced in STPA.
In our running example, we identified a list of 20 potential stakeholders. This includes directly involved entities such as end users, app developers, and API providers, as well as indirectly affected ones like trail management organizations and local businesses.\footnote{For the complete list of prompts and results in all steps, please refer to the supplementary material:
\url{https://docs.google.com/spreadsheets/d/1GkdN9TBscGhqfyFNdLMsLRtFKbr3DqqNzkaTDvW-XAQ/edit?usp=sharing}}
Our experiments show that LLMs can generate a list of stakeholders based on an application description.  We have designed prompts to encourage LLMs to consider indirect and less common stakeholders, thereby fostering a broad exploration. Developers can then use this list as inspiration for selecting stakeholders to analyze.

\begin{table}[t]
\caption{Values and losses among different stakeholders (Selected)}
\begin{center}
\begin{tabular}{p{.17\linewidth}p{.34\linewidth}p{.34\linewidth}}
\toprule
\textbf{Stakeholder} & \textbf{Value} & \textbf{Loss} \\ \midrule
{End User} & Personalization & Lack of personalization \\
& Accuracy of information & Inaccuracy of information \\ \addlinespace[2pt]

\multirow{2}{0.95\linewidth}{App Developers} & Maintainability & High maintenance burden \\ 
& Cost efficiency & Increased costs \\ \addlinespace[2pt]


\multirow{2}{0.95\linewidth}{Local Businesses} 
& Community engagement & Lack of community engagement \\ \bottomrule

\end{tabular}
\label{table: values and losses}
\end{center}
\end{table}

\textit{Losses:} For each stakeholder, we consider their values and convert them into corresponding losses. In our running example, we identified 145 losses, averaging 6-8 losses per stakeholder; a selection of these is shown in Table~\ref{table: values and losses}.
Again, we found LLMs helpful for suggesting values and losses, given the broad scope and the large number of losses they identified.

\textit{Hazards and Constraints:} For each loss, we identify hazards -- states or conditions that can lead to the loss under certain circumstances. We consider hazards beyond traditional STPA system boundaries. For example, the hazard ``system does not support user customization'' may lead to the loss ``lack of personalization''. Given a large number of losses and to avoid prematurely focusing on only a few, we find that LLMs effectively support analyzing all losses in a few minutes for under USD 2.
In our running example, we identified 1,159 hazards for 145 losses across 20 stakeholders, averaging 5-10 hazards per loss.
While some hazards are obvious, such as incorrect model outputs, we discovered many others we would not have anticipated without such systematic analysis, such as \textit{``System recommends trails in ecologically sensitive areas or during hazardous conditions.''}

\begin{table}[t]
\caption{System-level hazards (Selected)}
\begin{center}
\begin{tabular}{p{.5cm}p{7.5cm}}
\toprule
\textbf{HID} & \textbf{Hazard} \\ \midrule
H4 & System does not personalize recommendations effectively, support user customization, or adapt to user preferences and trends. \\ 
H24 & System recommends trails in ecologically sensitive areas or during hazardous conditions. \\ 
H39 & System lacks a centralized dashboard or knowledge repository for monitoring performance and user feedback. \\
H48 & System lacks a mechanism for users to withdraw consent or access their data records. \\ \bottomrule
\multicolumn{2}{l}{HID: Hazard ID}
\end{tabular}
\label{table: hazards}
\end{center}
\end{table}

At this stage, LLMs also effectively aid in merging similar hazards identified from different losses. In our running example, with human supervision for granularity, we reduced the hazards to 50 distinct ones, as examples shown in Table~\ref{table: hazards}.

\begin{table}[t]
\caption{Controller designs (Selected)}
\begin{center}
\begin{tabular}{p{0.035\linewidth}p{0.87\linewidth}}
\toprule
\textbf{HID} & \textbf{Controller Design}\\ \midrule

H4  & Integrate a feedback loop where users can provide feedback on trail recommendations, allowing the system to improve personalization. \\ \addlinespace[2pt]
 & Enable users to connect their social media or fitness tracking accounts for personalized recommendations. \\ 
\addlinespace



H39 & Develop a centralized dashboard that aggregates data from the LLM and external APIs to monitor system performance in real time.\\ \addlinespace[2pt]
  & Introduce automated alerts in the dashboard to notify administrators of any performance issues or anomalies detected in the system. \\ 
\addlinespace

H48  & Implement a user-friendly interface that allows users to easily withdraw consent for data collection and processing at any time. \\ \addlinespace[2pt]
 & Design the system to automatically notify users of any changes to data handling practices and obtain renewed consent if necessary.\\ 
\bottomrule
\multicolumn{2}{l}{HID: Hazard ID}
\end{tabular}
\label{table: controllers}
\end{center}
\end{table}

\begin{figure}[t]
\centerline{\includegraphics[width=\linewidth]{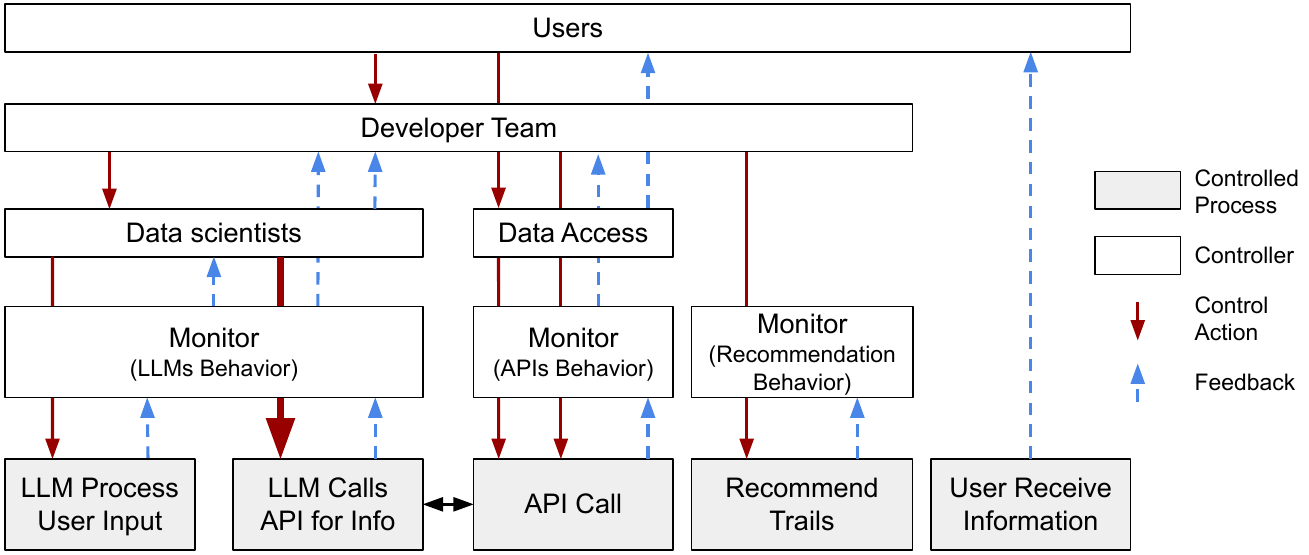}}
\caption{The control structure of the system: The bold arrow represents the control action analyzed in Section~\ref{section: analyzing and revising controllers}, involving data scientists upgrading or downgrading the LLM version or altering the prompts.}
\label{figure: Control Structure}
\end{figure}

\begin{table*}[t]
\caption{Unsafe control actions and loss scenarios (Selected)}
\begin{center}
\begin{tabular}{p{0.47\linewidth}p{0.47\linewidth}}
\toprule
\textbf{Unsafe Control Actions} & \textbf{Loss Scenario} \\ \midrule
Premature control actions, like upgrading the model without thorough analysis or confirmation, can destabilize operations or reduce user satisfaction. & Data scientists might overly rely on new model version improvements without properly assessing their impact on system interactions with APIs. \\ \addlinespace[2pt]
Delayed control actions can cause inefficiency, poor performance, user frustration, and a negative experience.
& Data scientists may face data overload, insufficient tools, or cognitive strain, hindering decision-making and delaying issues without clear prioritization.\\ \bottomrule
\end{tabular}
\label{table: UCA & Loss Scenarios}
\end{center}
\end{table*}

\subsection{Proactive Design of Control Structure}
Since the existing AllTrails application had few controllers, as is common in many ML-powered applications, there was not much to analyze regarding the existing control structures in STPA steps 2--4. However, the STPA process is also effective for \textit{proactively envisioning new controllers} to mitigate the identified hazards, aiding practitioners in \textit{modeling the control structure}, which is the second step in STPA.

We consider potential controllers or design structures that (a) prevent the hazard, (b) reduce its likelihood, (c) decrease the probability of it causing a loss, and (d) minimize the severity of the loss if it occurs.
In our running example, aided by LLMs, we identified 10 to 15 possible designs per hazard.
While some controllers may be impractical, costly, or insufficiently effective, our systematic approach revealed some easy-to-implement and potentially effective ideas we might not have otherwise considered, such as connecting social media or fitness tracking accounts for personalized recommendations.
Among the identified hazards, we chose to explore controllers for three hazards: H4, H39, and H48. These pertain to the lack of personalization, absence of system monitors, and inability for users to withdraw consent, respectively. Some potential controllers are listed in Table~\ref{table: controllers}. Having assessed the risks and costs, we adopted, modified, and merged several controller designs suggested by the LLMs. The resulting control structure incorporates human feedback and hierarchical monitoring mechanisms, which would be challenging to develop without this process. We then performed the second step of STPA and visually modeled the control structure of the system, as illustrated in Figure~\ref{figure: Control Structure}.

Again, we found that LLMs can help consider potential designs, especially when given instructions to explore different designs. However, developers remain crucial in making judgments about relevance and importance.

\subsection{Analyzing and Revising Controllers}~\label{section: analyzing and revising controllers}
After designing and modeling the control structure, we now return to steps 3 and 4 of the STPA process to analyze whether the controllers are sufficient or if they introduce new problems.

We identified \textit{unsafe control actions} by examining each control action against a checklist for potential constraint violations if the action was (a) absent, (b) incorrect, (c) mistimed, (d) executed in the wrong sequence, (e) prolonged, or (f) stopped early. We then identify potential \textit{loss scenarios}, which are states or events that may lead to the unsafe control action.

In our running example, we analyze the control action where data scientists modify the LLMs that call APIs, as depicted by the bold arrow in Figure~\ref{figure: Control Structure}.
Some identified unsafe control actions and corresponding loss scenarios are listed in Table~\ref{table: UCA & Loss Scenarios}. We found LLMs effective in systematically exploring different problems by following the checklist.

{
}


{
}



Practitioners can now review these unsafe control actions and loss scenarios to decide whether and how to revise certain controllers. For example, our initial monitoring approach was found to be naive, lacking an explicit structure to ensure that an operator regularly checks the dashboard, effectively detects issues, and does not cease monitoring due to notification fatigue. This encourages establishing explicit procedures for (a) setting up automated alerts, (b) implementing an engineering process for tracking and eliminating false positive alerts, and (c) designating someone responsible for monthly check-ins with the operator to ensure accountability. This approach goes far beyond simple engineering interventions such as exception handling, redundancy, or using an LLM to review the output of another LLM\cite{shankar2024validates}, instead considering the entire socio-technical system and its environment.

\section{Discussions and conclusion} \label{section: Discussion}

While others have advocated adopting safety engineering to anticipate harms in ML-powered applications \cite{khlaaf2022hazard, Dobbe_2022, martelaro2022exploring, rismani2023beyond, rismani2023plane, jatho2022system, rismani2024silos, rismani2021ai}, this paper reinforces this idea and provides a concrete example of the process, demonstrating its potential to anticipate harms beyond traditional safety concerns like bodily harm and mission loss.
Although our running example appears simple --merely combining a system prompt with an API on OpenAI's customized GPTs platform -- the analysis identified potential harms ranging from minor and far-fetched to worth addressing.
This encouraged comprehensive early system design to mitigate these harms.
We advocate that developers of novel ML-powered applications should undertake this practice, even if the application appears harmless. There is no justification for not attempting to anticipate the ``unintended consequences,'' given the uncertainty, potential bias, and possible shortcuts associated with ML components in software systems.

However, we also experienced firsthand how quickly the analysis can become unmanageable, especially when expanding beyond severe losses in traditional safety-critical systems. We analyzed dozens of stakeholders, hundreds of potential losses, and thousands of hazards. This scope can easily grow to include even more potential controllers and associated issues. Without technical support, this approach is overwhelming and can feel like a bureaucratic paper-heavy compliance activity, raising questions about how to encourage its routine adoption as part of everyday responsible engineering practices.

\paragraph*{Assisting Developers vs. Automating Compliance Activity} LLMs effectively assist in navigating STPA steps and extensively exploring stakeholders, losses, hazards, and controllers, helping to scale the process beyond just the severe losses. However, there is a risk of over-reliance on LLMs for fully automating hazard analysis. We envision LLMs as support tools, not replacements for developers' critical judgment. While LLMs can produce ideas and explore broadly by following checklists, developers are essential for thinking beyond suggestions, assessing severity, and setting the analysis focus. Balancing LLM support while preserving human creativity and judgment is a crucial question for future research.

\paragraph*{Expertise Requirements} Automated assistance in hazard analysis can enhance accessibility by offering step-by-step guidance and examples to developers lacking safety engineering training. However, there is a risk that these developers may perform a more superficial analysis. 
Ideally, increased accessibility enables selective engagement of safety experts, allowing their expertise to scale across more projects.
Balancing this requires further research, potentially drawing insights from similar discussions in software security \cite{Howard2003, Howard2006} and democratizing data science \cite{singh2022automated, drozdal2020trust}.

\paragraph*{Cataloging Common Controllers} ML research and much software engineering research on ML primarily focus on the model -- to improve accuracy, measure fairness, explain model internals, or rectify shortcuts. In contrast, system-level mitigations such as safeguards, user interface design, and human oversight receive less attention, despite their effectiveness as controllers \cite{Kastner2025}. Hazard analysis promotes broader system thinking, and we believe a pattern catalog of common system-level interventions could foster discussions and be embedded into the process of identifying and designing controllers.

\paragraph*{Fostering Adoption} Interview studies indicate practitioners hesitate to integrate hazard analysis into their agile and exploratory workflows, citing the need for organizational changes such as increased incentives, managerial confidence, support, understanding, and resource investment \cite{martelaro2022exploring, rismani2023plane}. Further research is required to embed hazard analysis into routine engineering, even in non-safety-critical contexts where success is a lack of issues. 
Lowering effort and demonstrating value, while appealing to developers' responsibility and mastery, may effectively change practices and culture. While much work remains, we can learn from past efforts such as DevOps\cite{luz2019adopting}, establishing a quality or security culture\cite{wiegers1996creating}, adopting fairness audits\cite{rakova2021responsible}, promoting social responsibility through education\cite{hironimus2009sociological}, and broadly altering organizational culture\cite{schein2010organizational}.
Technological innovations, education, and activism can each promote more responsible engineering practices.

\section*{Acknowledgment}

We thank Nadia Nahar, Laurie Williams, William Enck, and Alexandros Kapravelos for their feedback on this work.

\bibliographystyle{IEEEtran}
\bibliography{references}

\begin{thebibliography}{10}
\providecommand{\url}[1]{#1}
\csname url@samestyle\endcsname
\providecommand{\newblock}{\relax}
\providecommand{\bibinfo}[2]{#2}
\providecommand{\BIBentrySTDinterwordspacing}{\spaceskip=0pt\relax}
\providecommand{\BIBentryALTinterwordstretchfactor}{4}
\providecommand{\BIBentryALTinterwordspacing}{\spaceskip=\fontdimen2\font plus
\BIBentryALTinterwordstretchfactor\fontdimen3\font minus \fontdimen4\font\relax}
\providecommand{\BIBforeignlanguage}[2]{{%
\expandafter\ifx\csname l@#1\endcsname\relax
\typeout{** WARNING: IEEEtran.bst: No hyphenation pattern has been}%
\typeout{** loaded for the language `#1'. Using the pattern for}%
\typeout{** the default language instead.}%
\else
\language=\csname l@#1\endcsname
\fi
#2}}
\providecommand{\BIBdecl}{\relax}
\BIBdecl

\bibitem{bias_amplification_suresh2021framework}
H.~Suresh and J.~Guttag, ``A framework for understanding sources of harm throughout the machine learning life cycle,'' in \emph{Proceedings of the 1st ACM Conference on Equity and Access in Algorithms, Mechanisms, and Optimization}, 2021, pp. 1--9.

\bibitem{privacy_risks_chen2021machine}
M.~Chen, Z.~Zhang, T.~Wang, M.~Backes, M.~Humbert, and Y.~Zhang, ``When machine unlearning jeopardizes privacy,'' in \emph{Proceedings of the 2021 ACM SIGSAC conference on computer and communications security}, 2021, pp. 896--911.

\bibitem{pagano2022bias}
T.~P. Pagano, R.~B. Loureiro, F.~V.~N. Lisboa, G.~O.~R. Cruz, R.~M. Peixoto, G.~A. d.~S. Guimar{\~a}es, L.~L.~d. Santos, M.~M. Araujo, M.~Cruz, E.~L.~S. de~Oliveira \emph{et~al.}, ``Bias and unfairness in machine learning models: a systematic literature review,'' \emph{arXiv preprint arXiv:2202.08176}, 2022.

\bibitem{mehrabi2021survey}
N.~Mehrabi, F.~Morstatter, N.~Saxena, K.~Lerman, and A.~Galstyan, ``A survey on bias and fairness in machine learning,'' \emph{ACM computing surveys (CSUR)}, vol.~54, no.~6, pp. 1--35, 2021.

\bibitem{buccinca2023aha}
Z.~Bu{\c{c}}inca, C.~M. Pham, M.~Jakesch, M.~T. Ribeiro, A.~Olteanu, and S.~Amershi, ``Aha!: Facilitating ai impact assessment by generating examples of harms,'' \emph{arXiv preprint arXiv:2306.03280}, 2023.

\bibitem{wang2024farsight}
Z.~J. Wang, C.~Kulkarni, L.~Wilcox, M.~Terry, and M.~Madaio, ``Farsight: Fostering responsible ai awareness during ai application prototyping,'' in \emph{Proceedings of the CHI Conference on Human Factors in Computing Systems}, 2024, pp. 1--40.

\bibitem{microsoft_rai_template_2022}
\BIBentryALTinterwordspacing
Microsoft, ``Responsible {AI} impact assessment template,'' 2022, accessed: 2024-11-07. [Online]. Available: \url{https://blogs.microsoft.com/wp-content/uploads/prod/sites/5/2022/06/Microsoft-RAI-Impact-Assessment-Template.pdf}
\BIBentrySTDinterwordspacing

\bibitem{jatho2022system}
E.~W. Jatho~III, L.~O. Mailloux, S.~Rismani, E.~D. Williams, and J.~A. Kroll, ``System safety engineering for social and ethical ml risks: A case study,'' \emph{arXiv preprint arXiv:2211.04602}, 2022.

\bibitem{khlaaf2022hazard}
H.~Khlaaf, P.~Mishkin, J.~Achiam, G.~Krueger, and M.~Brundage, ``A hazard analysis framework for code synthesis large language models,'' \emph{arXiv preprint arXiv:2207.14157}, 2022.

\bibitem{Dobbe_2022}
\BIBentryALTinterwordspacing
R.~Dobbe, ``System safety and artificial intelligence,'' in \emph{2022 ACM Conference on Fairness, Accountability, and Transparency}, ser. FAccT ’22.\hskip 1em plus 0.5em minus 0.4em\relax ACM, Jun. 2022. [Online]. Available: \url{http://dx.doi.org/10.1145/3531146.3533215}
\BIBentrySTDinterwordspacing

\bibitem{martelaro2022exploring}
N.~Martelaro, C.~J. Smith, and T.~Zilovic, ``Exploring opportunities in usable hazard analysis processes for ai engineering,'' 2022.

\bibitem{rismani2023beyond}
S.~Rismani, R.~Shelby, A.~Smart, R.~Delos~Santos, A.~Moon, and N.~Rostamzadeh, ``Beyond the ml model: Applying safety engineering frameworks to text-to-image development,'' in \emph{Proceedings of the 2023 AAAI/ACM Conference on AI, Ethics, and Society}, 2023, pp. 70--83.

\bibitem{rismani2023plane}
S.~Rismani, R.~Shelby, A.~Smart, E.~Jatho, J.~Kroll, A.~Moon, and N.~Rostamzadeh, ``From plane crashes to algorithmic harm: applicability of safety engineering frameworks for responsible ml,'' in \emph{Proceedings of the 2023 CHI Conference on Human Factors in Computing Systems}, 2023, pp. 1--18.

\bibitem{rismani2024silos}
S.~Rismani, R.~Dobbe, and A.~Moon, ``From silos to systems: Process-oriented hazard analysis for ai systems,'' \emph{arXiv preprint arXiv:2410.22526}, 2024.

\bibitem{rismani2021ai}
S.~Rismani and A.~Moon, ``How do ai systems fail socially?: an engineering risk analysis approach,'' in \emph{2021 IEEE International Symposium on Ethics in Engineering, Science and Technology (ETHICS)}.\hskip 1em plus 0.5em minus 0.4em\relax IEEE, 2021, pp. 1--8.

\bibitem{adler2016safety}
R.~Adler, P.~Feth, and D.~Schneider, ``Safety engineering for autonomous vehicles,'' in \emph{2016 46th Annual IEEE/IFIP International Conference on Dependable Systems and Networks Workshop (DSN-W)}.\hskip 1em plus 0.5em minus 0.4em\relax IEEE, 2016, pp. 200--205.

\bibitem{abdulkhaleq2017using}
A.~Abdulkhaleq, S.~Wagner, D.~Lammering, H.~Boehmert, and P.~Blueher, ``Using stpa in compliance with iso 26262 for developing a safe architecture for fully automated vehicles,'' \emph{arXiv preprint arXiv:1703.03657}, 2017.

\bibitem{STPA_Handbook_Stpa2018}
{Leveson, Nancy G. and Thomas, John P.}, \emph{{STPA} handbook}, MIT Partnership for Systems Approaches to Safety and Security (PSASS), Cambridge, Massachusetts, U.S., 2018.

\bibitem{bender2021dangers}
E.~M. Bender, T.~Gebru, A.~McMillan-Major, and S.~Shmitchell, ``On the dangers of stochastic parrots: Can language models be too big?'' in \emph{Proceedings of the 2021 ACM conference on fairness, accountability, and transparency}, 2021, pp. 610--623.

\bibitem{weidinger2021ethical}
L.~Weidinger, J.~Mellor, M.~Rauh, C.~Griffin, J.~Uesato, P.-S. Huang, M.~Cheng, M.~Glaese, B.~Balle, A.~Kasirzadeh \emph{et~al.}, ``Ethical and social risks of harm from language models,'' \emph{arXiv preprint arXiv:2112.04359}, 2021.

\bibitem{zemel2013learning}
R.~Zemel, Y.~Wu, K.~Swersky, T.~Pitassi, and C.~Dwork, ``Learning fair representations,'' in \emph{International conference on machine learning}.\hskip 1em plus 0.5em minus 0.4em\relax PMLR, 2013, pp. 325--333.

\bibitem{shokri2015privacy}
R.~Shokri and V.~Shmatikov, ``Privacy-preserving deep learning,'' in \emph{Proceedings of the 22nd ACM SIGSAC conference on computer and communications security}, 2015, pp. 1310--1321.

\bibitem{MicrosoftAITools}
\BIBentryALTinterwordspacing
Microsoft, ``Responsible ai tools and practices,'' 2024, accessed: 2024-11-09. [Online]. Available: \url{https://www.microsoft.com/en-us/ai/tools-practices}
\BIBentrySTDinterwordspacing

\bibitem{kieslich2023anticipating}
K.~Kieslich, N.~Diakopoulos, and N.~Helberger, ``Anticipating impacts: Using large-scale scenario writing to explore diverse implications of generative ai in the news environment,'' 2023.

\bibitem{bogucka2024co}
E.~Bogucka, M.~Constantinides, S.~{\v{S}}{\'c}epanovi{\'c}, and D.~Quercia, ``Co-designing an ai impact assessment report template with ai practitioners and ai compliance experts,'' in \emph{Proceedings of the AAAI/ACM Conference on AI, Ethics, and Society}, vol.~7, 2024, pp. 168--180.

\bibitem{bahr2014system}
N.~J. Bahr, \emph{System Safety Engineering and Risk Assessment: A Practical Approach}, 2nd~ed.\hskip 1em plus 0.5em minus 0.4em\relax Boca Raton, FL: CRC Press, 2014.

\bibitem{leveson2004new}
N.~Leveson, ``A new accident model for engineering safer systems,'' \emph{Safety science}, vol.~42, no.~4, pp. 237--270, 2004.

\bibitem{SAE_ARP4761A_2023}
\BIBentryALTinterwordspacing
S.~International, ``Guidelines and methods for conducting the safety assessment process on civil airborne systems and equipment,'' SAE International, Tech. Rep. ARP4761A, December 2023. [Online]. Available: \url{https://www.sae.org/standards/content/arp4761a/}
\BIBentrySTDinterwordspacing

\bibitem{safeware_10.1145/202709}
N.~G. Leveson, \emph{Safeware: system safety and computers}.\hskip 1em plus 0.5em minus 0.4em\relax New York, NY, USA: Association for Computing Machinery, 1995.

\bibitem{safety_critical_computer_systems_10.5555/524721}
N.~R. Storey, \emph{Safety Critical Computer Systems}.\hskip 1em plus 0.5em minus 0.4em\relax USA: Addison-Wesley Longman Publishing Co., Inc., 1996.

\bibitem{Engineering_a_safer_world_Nancy2012}
N.~G. Leveson, \emph{{Engineering a Safer World: Systems Thinking Applied to Safety}}.\hskip 1em plus 0.5em minus 0.4em\relax The MIT Press, 01 2012.

\bibitem{GPT4_openai2023gpt4}
\BIBentryALTinterwordspacing
OpenAI, ``Gpt-4,'' 2023, accessed: 2024-11-04. [Online]. Available: \url{https://openai.com/research/gpt-4}
\BIBentrySTDinterwordspacing

\bibitem{chatgpt_alltrails}
------, ``Chatgpt assistant: Alltrails,'' \url{https://chatgpt.com/g/g-KpF6lTka3-alltrails}, accessed: 2024-10-25.

\bibitem{shankar2024validates}
S.~Shankar, J.~Zamfirescu-Pereira, B.~Hartmann, A.~Parameswaran, and I.~Arawjo, ``Who validates the validators? aligning llm-assisted evaluation of llm outputs with human preferences,'' in \emph{Proceedings of the 37th Annual ACM Symposium on User Interface Software and Technology}, 2024, pp. 1--14.

\bibitem{Howard2003}
M.~Howard and D.~LeBlanc, \emph{Writing Secure Code}, 2nd~ed.\hskip 1em plus 0.5em minus 0.4em\relax Redmond, WA: Microsoft Press, 2003.

\bibitem{Howard2006}
M.~Howard and S.~Lipner, \emph{The Security Development Lifecycle: SDL, a Process for Developing Demonstrably More Secure Software}.\hskip 1em plus 0.5em minus 0.4em\relax Redmond, WA: Microsoft Press, 2006.

\bibitem{singh2022automated}
V.~K. Singh and K.~Joshi, ``Automated machine learning (automl): an overview of opportunities for application and research,'' \emph{Journal of Information Technology Case and Application Research}, vol.~24, no.~2, pp. 75--85, 2022.

\bibitem{drozdal2020trust}
J.~Drozdal, J.~Weisz, D.~Wang, G.~Dass, B.~Yao, C.~Zhao, M.~Muller, L.~Ju, and H.~Su, ``Trust in automl: exploring information needs for establishing trust in automated machine learning systems,'' in \emph{Proceedings of the 25th international conference on intelligent user interfaces}, 2020, pp. 297--307.

\bibitem{Kastner2025}
C.~Kästner, \emph{Machine Learning in Production: From Models to Products}.\hskip 1em plus 0.5em minus 0.4em\relax Cambridge, MA: The MIT Press, 2025.

\bibitem{luz2019adopting}
W.~P. Luz, G.~Pinto, and R.~Bonif{\'a}cio, ``Adopting devops in the real world: A theory, a model, and a case study,'' \emph{Journal of Systems and Software}, vol. 157, p. 110384, 2019.

\bibitem{wiegers1996creating}
K.~E. Wiegers, \emph{Creating a software engineering culture}.\hskip 1em plus 0.5em minus 0.4em\relax Pearson Education, 1996.

\bibitem{rakova2021responsible}
B.~Rakova, J.~Yang, H.~Cramer, and R.~Chowdhury, ``Where responsible ai meets reality: Practitioner perspectives on enablers for shifting organizational practices,'' \emph{Proceedings of the ACM on Human-Computer Interaction}, vol.~5, no. CSCW1, pp. 1--23, 2021.

\bibitem{hironimus2009sociological}
R.~J. Hironimus-Wendt and L.~E. Wallace, ``The sociological imagination and social responsibility,'' \emph{Teaching Sociology}, vol.~37, no.~1, pp. 76--88, 2009.

\bibitem{schein2010organizational}
E.~H. Schein, \emph{Organizational culture and leadership}.\hskip 1em plus 0.5em minus 0.4em\relax John Wiley \& Sons, 2010, vol.~2.

\end{thebibliography}

\end{document}